\documentclass[%
aps,prb,preprint,
superscriptaddress,
nobibnotes,
amsmath,amssymb,
floatfix,
]{revtex4-2}

\usepackage{xcolor}
\usepackage{graphicx}
\usepackage{dcolumn}
\usepackage{bm}

\usepackage{caption}
\begin{document}

\title{Thermal Conductivity of Water at Extreme Conditions}
\author{Cunzhi Zhang}
    \thanks{These two authors contributed equally.}
    \affiliation{Pritzker School of Molecular Engineering, University of Chicago, Chicago, IL 60637, USA}
\author{Marcello Puligheddu}
    \thanks{These two authors contributed equally.}
    \affiliation{Pritzker School of Molecular Engineering, University of Chicago, Chicago, IL 60637, USA}
    \affiliation{Materials Science Division and Center for Molecular Engineering, Argonne National Laboratory, Lemont, IL 60439, USA}
\author{Linfeng Zhang}
    \affiliation{Program in Applied and Computational Mathematics, Princeton University, Princeton, NJ 08544, USA}
\author{Roberto Car}
    \affiliation{Program in Applied and Computational Mathematics, Princeton University, Princeton, NJ 08544, USA}
    \affiliation{Department of Chemistry, Department of Physics, and Princeton Institute for the Science and Technology of Materials, Princeton University, Princeton, NJ 08544, USA}
\author{Giulia Galli}
\email{gagalli@uchicago.edu}
    \affiliation{Pritzker School of Molecular Engineering, University of Chicago, Chicago, IL 60637, USA}
    \affiliation{Materials Science Division and Center for Molecular Engineering, Argonne National Laboratory, Lemont, IL 60439, USA}
    \affiliation{Department of Chemistry, University of Chicago, Chicago, IL 60637, USA}

\date{\today}

\begin{abstract}

Measuring the thermal conductivity ($\kappa$) of water at extreme conditions is a challenging task and few experimental data are available. We predict $\kappa$ for temperatures and pressures relevant to the conditions of the Earth mantle, between  1,000 and 2,000 K and up to 22 GPa. We employ close to equilibrium molecular dynamics  simulations and a deep neural network potential fitted to density functional theory data. We then interpret
our results by computing the equation of state of water on a fine grid of points and using a
simple model for $\kappa$. We find that the thermal conductivity is weakly dependent on temperature and monotonically increases
with pressure with an approximate square-root behavior. In addition we show how the increase of $\kappa$ at high pressure, relative to ambient conditions, is related to the corresponding
increase in the sound velocity. Although the relationships between the thermal conductivity, pressure and sound velocity established here are not rigorous, they are sufficiently accurate to allow for a robust estimate of the thermal conductivity of water in a  broad range of temperatures and pressures, where experiments are still difficult to perform.


\end{abstract}


\maketitle
\clearpage

\section{Introduction}

Water under pressure ($P$)  at high temperature ($T$) is an important constituent of the continental crust of the Earth \cite{manning2018Ann_water_crust}, and of the interiors of ice giants, e.g. Uranus and Neptune \cite{nettel2013PSS_water_interior}, as well as many exoplanets \cite{zeng2019PNAS_water_interior,baraffe2010RPP_water}.
Hence the characterization of the heat transport properties of water at extreme conditions is central to Earth and planetary sciences \cite{guillot2005AREPS_planetary,fortney2010SSR_planetary,schubert2011PEPI_planetary}. 
For example, understanding water heat transport may help explain  the remarkably low luminosity of Uranus \cite{nettel2016Ica_luminosity} as well as derive models for the core erosion processes in Jupiter \cite{moll2017Ast_erosion}.
However, the thermal conductivity ($\kappa$) of water at high $P$ and $T$ (HPT)  is  poorly known and difficult to measure.

Direct measurements at extreme conditions are challenging not only because of the reactivity of water, but also for the errors that may be introduced in the experiments by  convection and radiation processes \cite{Ross1984RPP_TC_measure}. 
Measurements for liquid water are available for $P <$ $\sim$ 3.5 GPa and $T <$ $\sim$ 1,000 K \cite{huber2012JPCRD_TC_data,abramson2001JCP_TC_3.5GPa} and for ice up to 22 GPa \cite{chen2011PRB_TC_22GPa}, below  1,000 K.

Similar to experiments, simulations of heat transport in water at HPT  are challenging. Reliable empirical force-fields are not available and so far first principles molecular dynamics (FPMD) simulations based on density functional theory (DFT) have been mostly limited to structural, vibrational and electronic properties \cite{zhao2020PCCP_AIMD_EOS, rozsa2018PNAS_AIMD_Conductivity, zhang2020NC_AIMD_Salt,pan2014NC_AIMD_Electronic,french2011PRL_AIMD_Ionic,goncharov2005PRL_AIMD_Ionization}, due to the long simulation times and large unit cells usually required to investigate transport properties. However, recently French \cite{french2019NJP_TC_Calc} conducted FPMD simulations of water at extreme conditions to obtain its thermal conductivity, although the heat current was approximated by fitting pair potentials. 
In addition, thanks to important theoretical advances \cite{marco2016NP_TC_flux,ercole2016JLTP_TC_gauge}, ab initio calculations of the thermal conductivity of water using linear response and the Green-Kubo (GK) formalism  \cite{green1952JCP_GK,green1954JCP_GK,kubo1957JPSJ_GK_I,kubo1957JPSJ_GK_II} have been conducted both at ambient \cite{marco2016NP_TC_flux} and extreme conditions \cite{grasselliNC_TC_Calc}, but only for pressures higher than 33 GPa.
However, analytical expressions for the energy density and flux required in GK calculations are not be easily available for sophisticated DFT functionals \cite{tisi2021PRB_TC_Calc}; furthermore,  despite  novel noise-reduction methods \cite{ercole2017SR_cepstral,bertossa2019PRL_cepstral}, long simulation times of several hundreds of pico-seconds for systems with several hundred of atoms are  required to obtain converged results for the thermal conductivity, making first principles simulations a rather demanding task.
Hence a computational framework avoiding the explicit calculation of the heat flux and allowing for long simulation times is desirable, to explore the thermal conductivity of water in a wide range of conditions.


Here we use the sinusoidal approach to equilibrium molecular dynamic (SAEMD)
method recently proposed for fluids \cite{puligheddu2020PRM_SAEMD} with a   deep neural network potential (DP) \cite{wang2018CPC_deepmd,zhang2018PRL_deepmd,zhang2018ANIPS_deepmd}, allowing for long simulation times with relatively large cells. 
The DP inter-atomic potential, trained on first-principles data, can accurately describe interatomic interactions at a cost slightly higher than that of classical force-fields, but much lower than FPMD. 
We compute the thermal conductivity of water for 1,000 $< T <$ 2,000 K and 1.0 $< \rho <$ 1.86 g/cm$^3$, namely at conditions relevant to the Earth mantle.
We then interpret our results by computing the equation of state (EOS) of water on a fine grid of points and using a simple model derived from our EOS results and the computed values of $\kappa$. We find that at the conditions studied here $\kappa$ increases relative to ambient conditions, is weakly dependent on temperature and monotonically increases with pressure with an approximate square-root behavior.

The rest of the paper is organized as follows. The methods used here to compute the thermal conductivity and equation of state are described in the next section, followed by a presentation of our results and finally by our conclusions.

\section{Methods}
\subsection{Thermal conductivity calculations}
We investigated the thermal conductivity of water at high pressure and temperature by carrying  out molecular dynamics simulations with a deep neural network potential  \cite{wang2018CPC_deepmd,zhang2018PRL_deepmd,zhang2018ANIPS_deepmd} and the LAMMPS code \cite{plimpton1995JCP_lammps,thompson2022CPC_lammps}.
The potential was trained with the DeepMD-kit package \cite{wang2018CPC_deepmd} using ice and water structures from low temperature and pressure to about 2,400 K and 50 GPa. 
The training data were obtained from density functional theory calculations using the strongly constrained and appropriately normed
(SCAN) meta-GGA exchange correlation functional \cite{sun2015PRL_SCAN}. 
More details can be found in Ref. \cite{zhang2021PRL_phase}.

Specifically, we used the SAEMD method \cite{puligheddu2020PRM_SAEMD}, which allowed us to avoid the calculation of the heat flux, and we computed the thermal conductivity of the liquid from its response to a perturbation. 
This perturbation is a non-homogeneous constant temperature profile $T(x,y,z)$, maintained by a thermostat:
\begin{equation}
    T(x,y,z) = T_0 + \frac{\Delta T}{8} \Big(
        \Big(1-\cos(\frac{2\pi x}{L})\Big)
        \Big(1-\cos(\frac{2\pi y}{L})\Big)
        \Big(1-\cos(\frac{2\pi z}{L})\Big) - 4 \Big)
\end{equation}
where $L$ is the length of the simulation cell chosen to represent the system, and $\Delta T$ is the difference between the maximum and the minimum temperature. During  MD simulations we monitored how much energy the thermostat is providing to the system, and computed the thermal conductivity from the solution of the heat equation:
\begin{equation}
\label{eq:steady}
    0 = \kappa \nabla^2 T + q
\end{equation}
where $q$ is the heat generation rate per unit volume from the thermostat.

We carried out eight simulations: (i) one at ambient conditions, at $T=$ 300 K and $\rho=$ 1 g/cm$^{3}$; (ii) three calculations at $T=$ 1,000 K and $\rho$ $\in$ [1.2, 1.57, 1.86] g/cm$^{3}$; (iii) four calculations at $T=$ 2,000 K and $\rho$ $\in$ [1.0, 1.2, 1.57, 1.86] g/cm$^{3}$. We do not report calculations for $\rho=$ 1 g/cm$^{3}$ and $T=$ 1,000 K as it was difficult to properly converge our simulations due to the presence of large fluctuations in the heat generation rate $q$.
We used $\Delta T =$ 10, 30 and 100 K for calculations at $T=$ 300, 1,000 and 2,000 K respectively.
For each combination of density and temperature, we performed 20 independent runs, over which we averaged the amount of energy transferred to the system to compute the thermal conductivity.  
We used a cubic cell containing 512 water molecules, which was large enough to obtain approximately converged results, as previously verified \cite{puligheddu2020PRM_SAEMD}.
For example, at $T$ = 1,000 K and $\rho$ = 1.57 g/cm$^{3}$ SAEMD simulations with 512 molecules  yield a slight underestimation of the thermal conductivity of $\sim$ 5 \%, compared to the extrapolated value to infinite size. 

For each independent run, we equilibrated the system for $3\times10^{5}$ steps, followed by production runs of $10\times10^{5}$ steps.
We used a time-step of 0.2 fs and collected data for a total of 4 ns  for $\rho=$ 1.0 and 1.2 g/cm$^{3}$, and we used a time-step of 0.25 fs and collected data for a total of 5 ns for $\rho=$ 1.57 and 1.86 g/cm$^{3}$.

\subsection{Equation of state calculations}
We also carried out equation of state calculations by considering 90 $T$-$\rho$ conditions on an evenly spaced 9 $\times$ 10 mesh, for 1,000 $< T <$ 2,000 K (9 grid points) and 1.0 $< \rho <$ 1.9 g/cm$^{3}$ (10 grid points). 
At each $T$-$\rho$ condition, we performed MD simulations using the DP potential in the NVT ensemble with a time-step of 0.2 fs and a cubic cell containing 128 water molecules.
For each MD simulation, we equilibrated the system for 20 ps, followed by a production run of 54 ps.
In order to test finite size effects, we compared total energies and pressures obtained when using  cubic cells of 128 and 512 water molecules at $\rho$ = 1.2 g/cm$^{3}$ and $T$ from 1,000 to 2,000 K; the relative differences in total energy are $<$ 0.1 \% and those in pressure are $<$ 1 \%, which are attributed to statistical errors.
However, at $T=$ 1,000 K and $\rho=$ 1.8 and 1.9 g/cm$^{3}$ we found that the system did not exhibit a diffusive behavior, when using 128 water molecules in our cell. Hence we discarded the results of these simulations and  we used a larger  cell (512 water molecules) at $T=$ 1,000 K and $\rho=$ 1.86 g/cm$^{3}$, where the system  did behave as a liquid; for this simulation we used a time-step of 0.2 fs and equilibrated the system for 30 ps, followed by a  production run of 120 ps.

At each $T$-$\rho$ condition we computed the total energy ($E$), the pressure ($P$) and the water dissociation ratio,  obtained by  using a cutoff distance for O–H bonds of 1.25 Å. 
We then interpolated  $E(T,\rho)$, $P(T,\rho)$ and the water dissociation ratio over the whole parameter range considered here, by using the Gaussian process regression method as implemented in the sklearn package \cite{scikit-learn}. 
We used the radial basis function kernel with independent length-scales for $T$ and $\rho$.
The hyper-parameters of the model were obtained by maximizing the log-marginal-likelihood.

Based on the interpolated functions, which are differentiable, we computed  additional  properties of the system.
In particular, we computed the constant volume heat capacity per atom ($C_{V}$) as:
\begin{equation}
\label{eq:CV}
C_{V} (T, \rho) = \Bigl( \frac{ \partial E(T, \rho) }{ \partial T} \Bigr)_{\rho} 
\end{equation}
where $E(T, \rho)$ here is the total energy per atom. Further, we computed the constant pressure heat capacity per atom ($C_{P}$) \cite{cheng2020PRL_CP_CV} as:
\begin{equation}
\label{eq:CP}
C_{P} (T, \rho) = C_{V}(T, \rho) + 
           \frac{ T m }{ \rho^2 } 
    \Bigl( \frac{ \partial P(T, \rho) }{ \partial T }    \Bigr)_{\rho}^{2} 
    \Bigl( \frac{ \partial P(T, \rho) }{ \partial \rho } \Bigr)_{T}^{-1}
\end{equation}
where $m$ is the average mass per atom. 
We also obtained the adiabatic index as $\gamma (T, \rho) = C_{P} (T, \rho) / C_{V}(T, \rho)$, and  computed the sound velocity $C_{S}$ \cite{alavi1995science_CS} as:
\begin{equation}
\label{eq:CS}
C_{S} (T, \rho) = \sqrt{ \gamma (T, \rho) } 
   \sqrt{ \Bigl( \frac{ \partial P(T, \rho) }{ \partial \rho } \Bigr)_{T} }
\end{equation}

We calculated all the properties described above on a dense 100 $\times$ 40 mesh, for 1,000 $< T <$ 2,000 K (100 point grid) and 1.0 $< \rho <$ 1.9 g/cm$^{3}$ (40 point grid).

\section{Results and discussion}

\subsection{Computed thermal conductivity}

\begin{figure*}[!htp]
\centering
  \includegraphics[width=0.5\textwidth]{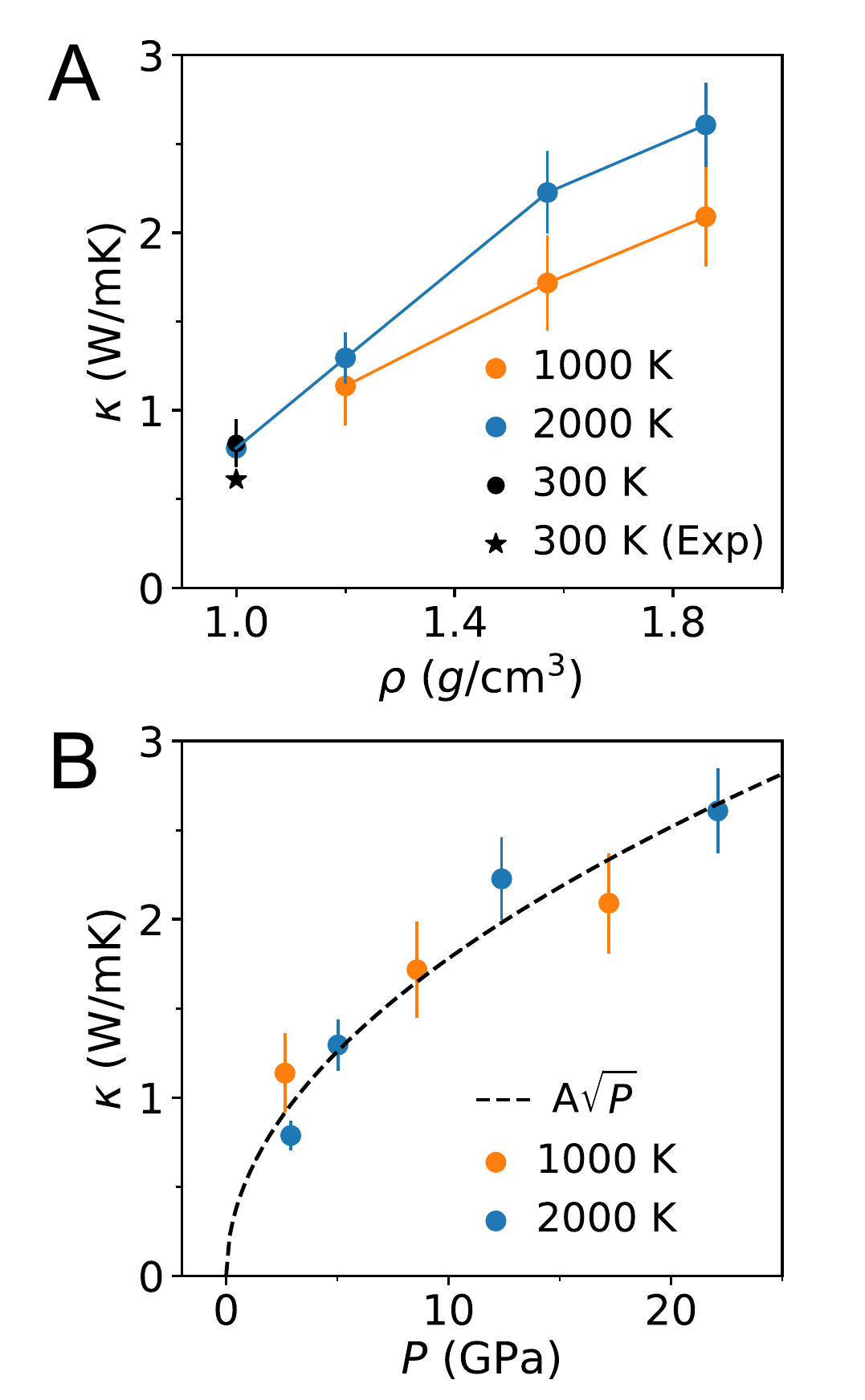}
  \caption{Thermal conductivity $\kappa$ of water at extreme conditions. A: dependence of $\kappa$ on density ($\rho$). We also show $\kappa$ at ambient conditions, i.e. the measured (300 K (Exp)) and computed (300 K) values. B: dependence of $\kappa$ on pressure ($P$). The dashed line is a simple fit $\kappa = A\sqrt{P}$ ($A\approx$ 0.56).  }
  \label{fig:kappa_rho_P}
\end{figure*}

Our computed values of the thermal conductivity $\kappa$ at extreme conditions are summarized in Table \ref{tab:kappa}. We also present results at ambient conditions for comparison. 
In Figure \ref{fig:kappa_rho_P}, we show  $\kappa$ of water at extreme conditions as a function of the density (Figure \ref{fig:kappa_rho_P}A) and  pressure (Figure \ref{fig:kappa_rho_P}B).

\begin{table}
\caption{Thermal conductivity ($\kappa$) of water at ambient and extreme conditions as obtained from SAEMD simulations using the DP potential. We also report the standard deviation error $\Delta \kappa$.}
\setlength\extrarowheight{-3pt}
\begin{tabular}{| c | c | c | c | c |}
 \hline
  $\,$ Temperature (K)        $\,$ & 
  $\,$ Density (g/cm$^3$)     $\,$ &
  $\,$ Pressure (GPa)         $\,$ &
  $\,$ $\kappa$ (W/mK)        $\,$ &
  $\,$ $\Delta \kappa$ (W/mK) $\,$ \\
 \hline
300  &  1.00 &   10$^{-4}$ &  0.81 &  0.14 \\
1000 &  1.20 &   2.6 &  1.14 &  0.22 \\
1000 &  1.57 &   8.6 &  1.72 &  0.27 \\
1000 &  1.86 &  17.2 &  2.09 &  0.28 \\
2000 &  1.00 &   2.9 &  0.79 &  0.08 \\
2000 &  1.20 &   5.0 &  1.29 &  0.15 \\
2000 &  1.57 &  12.4 &  2.23 &  0.23 \\
2000 &  1.86 &  22.1 &  2.61 &  0.24 \\
\hline
\end{tabular}
\label{tab:kappa}
\end{table}

We start by comparing our results at ambient conditions to those of previous studies and experiments. The calculated value of 0.81 W/mK at  300 K and 1 g/cm$^3$, agrees relatively well with that obtained via spectral analysis of the energy flux in NVE simulations with 128 water molecule cells\cite{tisi2021PRB_TC_Calc}, as expected  since both studies used the DP potential trained on a SCAN-generated data set. Based on the finite-size scaling study reported in Ref. \cite{puligheddu2020PRM_SAEMD} using empirical potentials, we expect our results to represent an underestimate of the data one would obtain for infinite sizes (possibly up to 15$\%$). The overall over-estimate from simulations compared to the experimental value (0.609 W/mK \cite{Powell1966_TC_exp,Ramires1995_TC_exp})  may be due to the neglect of nuclear quantum effects and  to errors introduced by the SCAN functional. We note that when using the DP model at the SCAN level of theory, the freezing temperature of water is $\sim$ 310 K \cite{piaggi2021JCTC_freez_310}. 
At 300 K, water described by the SCAN functional is sluggish and solid-like; hence it is not surprising that the thermal conductivity at 300 K is overestimated with this functional. 
Nuclear quantum effects are known to considerably affect the internal vibrations of water molecules\cite{ceriotti2016CR_ZPE}; however, the contribution to heat conduction of intra-molecular motion is probably not substantial at ambient conditions, where the major contributions are expected to come from low frequency modes. However the effect of the quantum nuclear motion on the thermal conductivity remains to be established and will be the topic of a future investigation.

We now turn to analyzing the dependence of $\kappa$ on  the temperature, density (Figure \ref{fig:kappa_rho_P}A) and pressure (Figure \ref{fig:kappa_rho_P}B).
At the densities studied here, we find that the thermal conductivity increases slightly with $T$ from 1,000 to 2,000 K.
Incidentally, the $\kappa$ for water at 1 g/cm$^3$ and 2,000 K is  almost the same as that computed at ambient conditions.
Consistent with experimental data at lower temperature and pressure \cite{abramson2001JCP_TC_3.5GPa}, and with high pressure studies of ice \cite{chen2011PRB_TC_22GPa}, our simulations show an increase of the thermal conductivity as the density and pressure are increased. 
In addition, our results are consistent with the simulation reported in Ref. \cite{french2019NJP_TC_Calc}, where the authors found that the thermal conductivity in the $T$-$\rho$ range investigated in our work is approximately  independent on temperature.
Remarkably, we find that a square root function $\kappa = A\sqrt{P}$  captures rather well the dependence of $\kappa$ on $P$, at both 1,000 and 2,000 K ($A$ is a  parameter almost constant as a function of $T$, between 1,000 and 2,000 K).
We quantify the relative error (RE) as:
\begin{equation}
\label{eq:re_fit}
{\rm RE} = \frac{\lvert A\sqrt{P} - \kappa \rvert} { \kappa } \times 100\%
\end{equation}
where the fitted $A$ ($\approx$ 0.56) was used, and $\kappa$ is the thermal conductivity from SAEMD simulations. We find the average RE over the 7 data points is 10.3 \%; the maximum RE is 21.9 \%. 

In order to interpret the temperature, pressure and density dependence of $\kappa$, particularly the relation $\kappa = A\sqrt{P}$ found in our simulation, we employed a simple model, described below.

\subsection{Model to interpret simulations}

\begin{figure*}[!htp]
\centering
  \includegraphics[width=0.99\textwidth]{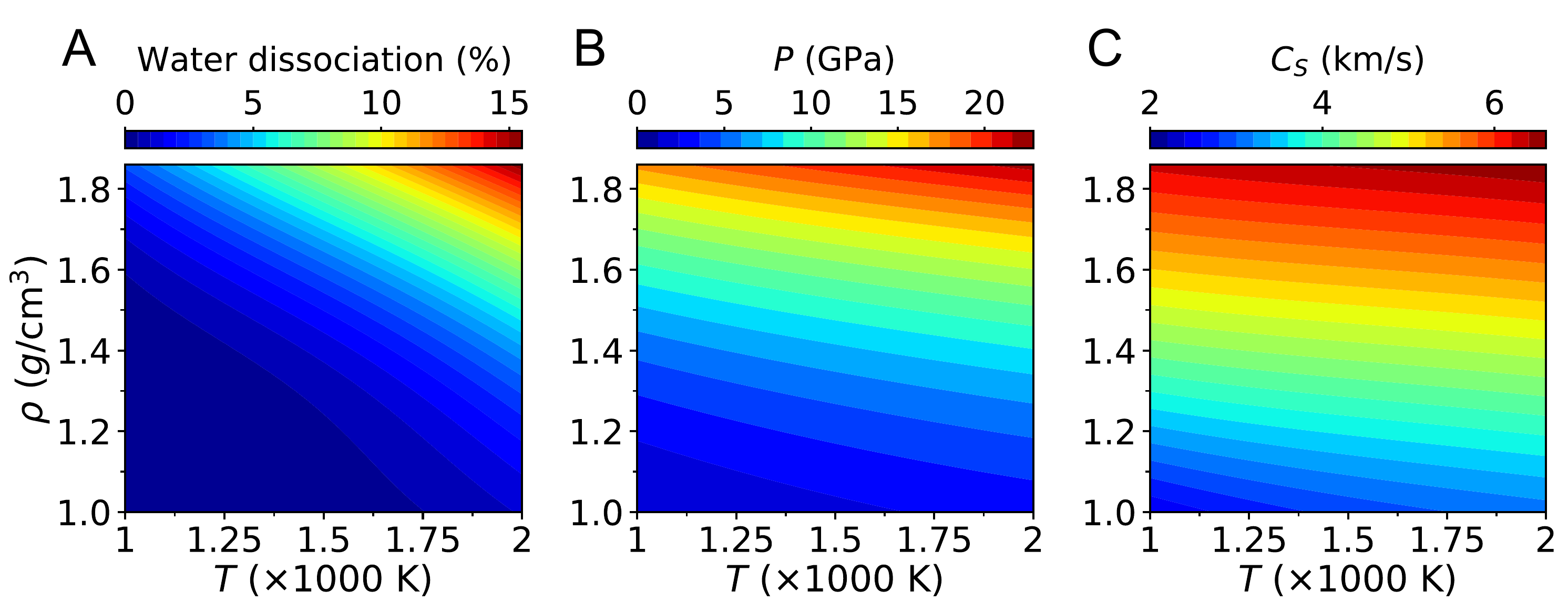}
  \caption{ Interpolated physical properties of water at extreme conditions. A: the ratio of dissociated molecules; B: the pressure ($P$); C: the sound velocity ($C_{S}$). }
  \label{fig:EOS_HPT}
\end{figure*}

\begin{figure*}[!htp]
\centering
  \includegraphics[width=0.6\textwidth]{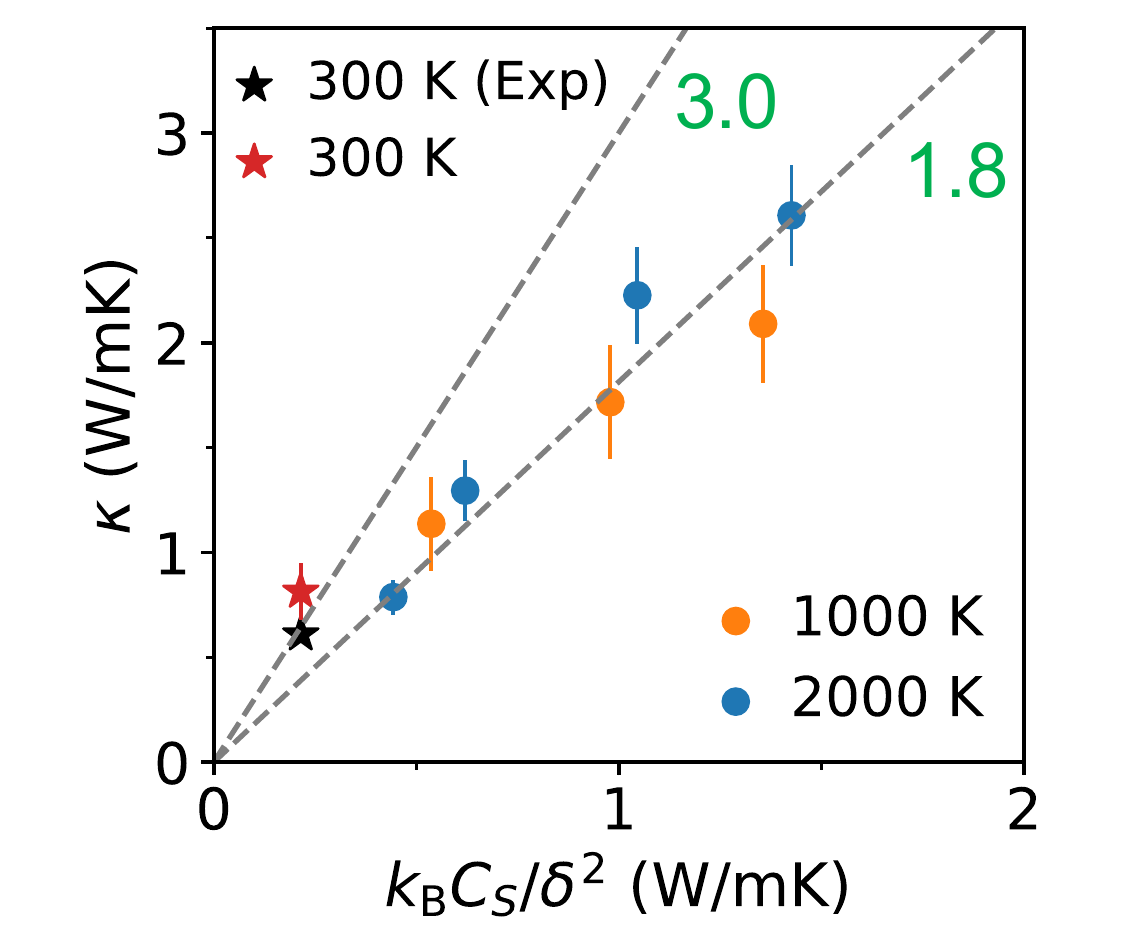}
  \caption{ Validation of the model
   $\kappa = \widetilde{C_{V}} k_{\rm B}C_{S}/\delta^{2}$ (eq \ref{eq:kappa}) used to interpret simulation data, where  $\widetilde{C_{V}}$ is a fitting parameter. We  show a dashed line for  $\widetilde{C_{V}}\approx1.8$, corresponding to the best fit of the values of $\kappa$ at extreme conditions. 
   The measured (300 K (Exp)) and computed (300 K) values at ambient conditions are also shown using the measured sound velocity of water $\approx$ 1500 m/s, as well as a dashed line for $\widetilde{C_{V}}=3.0$.
   }
  \label{fig:model_fit}
\end{figure*}

Numerous models have been proposed in the literature to describe  the thermal conductivity of liquids \cite{Bridgman1923PAAAS_TC_liquid,khrapak2021PRE_TC_liquid,zhao2021JAP_TC_liquid,chen2022JHT_TC_liquid}. Here we use a simple expression of $\kappa$  encompassing several of these models:
\begin{equation}
\label{eq:kappa}
\kappa = \widetilde{C_{V}} k_{\rm B}C_{S}/\delta^{2}
\end{equation}
where $\widetilde{C_{V}}$ is a model-dependent prefactor; $k_{\rm B}$ is the Boltzmann constant; $C_{S}$ is the sound velocity; $\delta = n^{-1/3}$ is the inter-molecular distance, where $n$ is the number density of the molecules in the fluid. 
In eq \ref{eq:kappa} one assumes that the amount of energy transferred during heat transport is proportional to $\widetilde{C_{V}} k_{\rm B}$ and that the speed of energy transfer is approximately equal to the sound velocity $C_{S}$; the energy is transferred step by step, between neighboring molecules separated by a distance  $\delta$.

We extend the use of eq \ref{eq:kappa} to interpret the results of  the thermal conductivity of water computed at extreme conditions. 
It should be noted that at HPT water may dissociate\cite{ rozsa2018PNAS_AIMD_Conductivity, zhang2020NC_AIMD_Salt,ye2022FD_AIMD_PE,french2011PRL_AIMD_Ionic,goncharov2005PRL_AIMD_Ionization,zhang2021PRL_phase}. Hence, we first verified whether the use of eq \ref{eq:kappa}, derived for simple liquids with no dissociating units, is at least approximately justified. 
We computed the ratio of dissociated water molecules in our samples, as shown in Figure \ref{fig:EOS_HPT}A.
We found that even at the highest $T$ and $\rho$ studied here, less than  $\sim$ 15 \%  of molecules were dissociated and therefore we expect that  the dissociation of water molecules is not a major factor affecting thermal transport at the conditions considered in our work.
Hence the use of the model of eq \ref{eq:kappa} appears to be reasonable to  interpret our simulation results for HPT water.

Similar to previous studies, we use $\delta = (\rho/m_{0})^{-1/3}$, where $m_{0}$ is the mass of a water molecule. Using $\delta$ equal to the average O-O distance in water yields similar results.
The value of $\widetilde{C_{V}}(T, \rho)$ and $C_{S} (T, \rho)$, the sound velocity of water at extreme conditions, are not available from the literature.
We can obtain the $C_{S} (T, \rho)$ from our equation of state calculations for $E(T, \rho)$ and $P(T, \rho)$ (see Methods), shown in Figure 2B and 2C.
However, we do not have well-defined methods to compute $\widetilde{C_{V}}(T, \rho)$, especially at HPT.
In previous studies,  $\widetilde{C_{V}}$ was approximated by the specific heat per molecule ($C_{V}$ or $C_{P}$) \cite{khrapak2021PRE_TC_liquid,zhao2021JAP_TC_liquid,chen2022JHT_TC_liquid}, e.g. $\widetilde{C_{V}} = C_{V} / k_{\rm B}$. 
This may be a good approximation for liquids, including water, at near ambient conditions, where the major contribution to the heat capacity comes from inter-molecular interactions.
However, at extreme conditions and high temperature the contributions of intra-molecular vibrations cannot be ignored.
Therefore, we would expect a serious error in our estimate of $\kappa$ if we used $\widetilde{C_{V}}$ as $C_{V} / k_{\rm B}$ in eq \ref{eq:kappa}. 
Hence here we  treat $\widetilde{C_{V}}$ as a parameter that we fit using the computed $\kappa$ at high $P$ and $T$ (the 7 data points in Table \ref{tab:kappa}).

As shown in Figure \ref{fig:model_fit}, we obtain a reasonable linear fit of $\kappa$ versus $k_{\rm B}C_{S}/\delta^{2}$, from which we determine $\widetilde{C_{V}}$ $\approx$ 1.8. 
We quantify the RE as:
\begin{equation}
\label{eq:RE_model}
{\rm RE} = \frac{\lvert \widetilde{C_{V}} k_{\rm B}C_{S}/\delta^{2} - \kappa \rvert} { \kappa } \times 100\%
\end{equation}
where the fitted $\widetilde{C_{V}}$ ($\approx$ 1.8) was used, and $\kappa$ is the thermal conductivity from SAEMD simulations. We find  that the average RE over the 7 data points  is 9.5 \%; the maximum RE is 17.7 \%. 
The reasonable error found here indicates that water dissociation is unlikely to  affect heat conduction in HPT water, in the $T$-$\rho$ range considered in  our work. However, while dissociation remains limited, proton conduction via Grotthus like mechanisms might play a role in determining heat transport. This aspect has not been studied in detail here and also for this reason we chose to fit the $\widetilde{C_{V}}$ parameter to simulation data.  
To show qualitatively  the difference between $\widetilde{C_{V}}$ at ambient and extreme conditions,
in Figure \ref{fig:model_fit} we plot two  lines corresponding to a value of  $\widetilde{C_{V}}$ equal to 3 and  $\approx$ 1.8. 
When using $\widetilde{C_{V}}=3$ \cite{xi2020CPL_TC_liquid,khrapak2021PRE_TC_liquid}, the measured value of $\kappa$ at ambient conditions can be correctly predicted.
The smaller $\widetilde{C_{V}}$ ($\approx$ 1.8) found at HPT appears to be consistent with the presence of a disrupted hydrogen bonded network and a small fraction of dissociated water molecules at extreme conditions,  leading to a  decrease in the energy transfer between adjacent molecules, relative to ambient conditions. 

We expect that treating $\widetilde{C_{V}}$ as a function of $T$ and $\rho$, instead of  a fitting parameter (Figure \ref{fig:model_fit}), would increase the accuracy of the model (eq \ref{eq:kappa}) in describing the thermal conductivity at HPT.

\begin{figure*}[!htp]
\centering
  \includegraphics[width=0.6\textwidth]{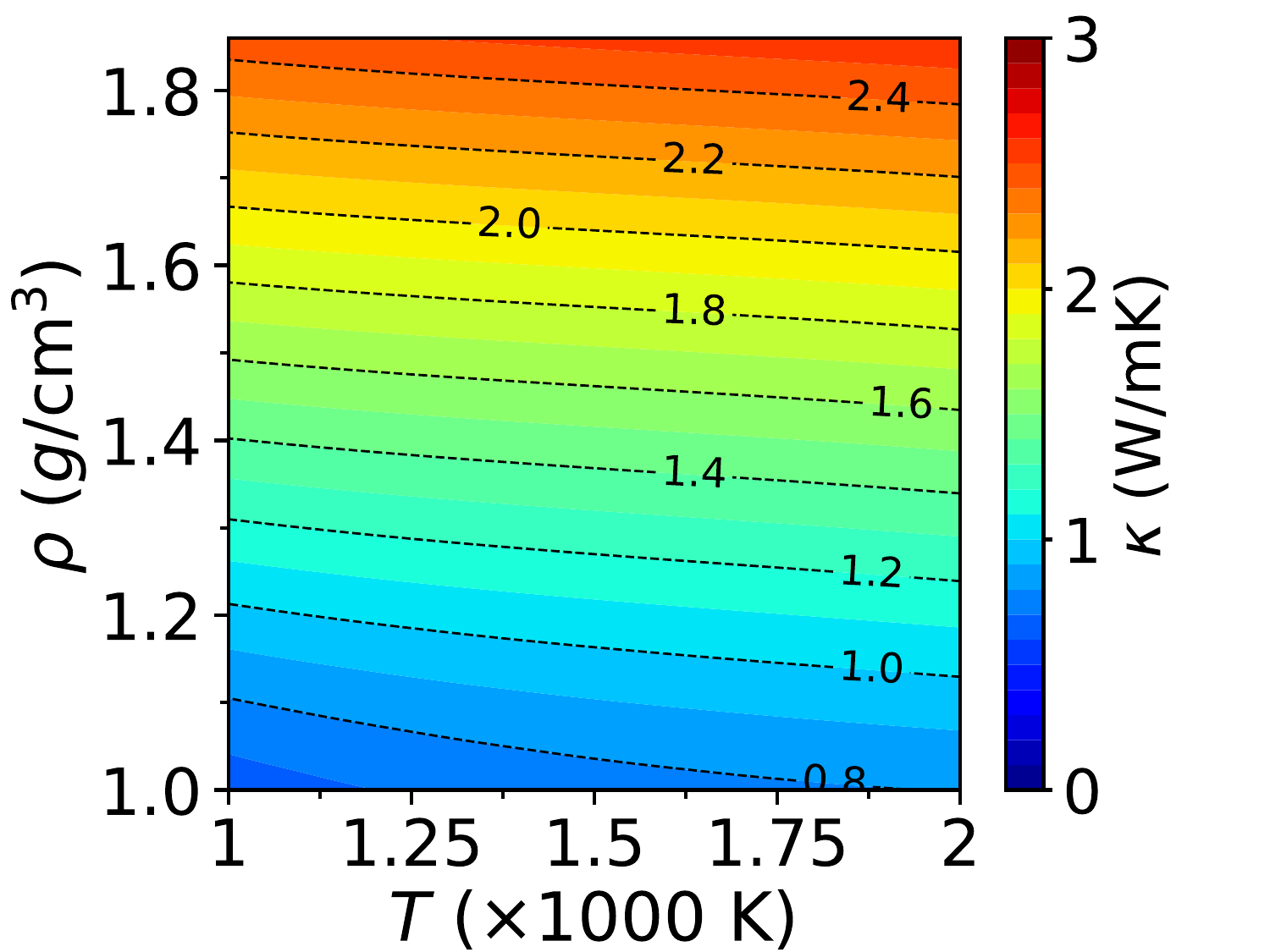}
  \caption{ Predicted thermal conductivity $\kappa$ using the model (eq \ref{eq:kappa}; see text).}
  \label{fig:kappa_pred}
\end{figure*}

Using the model (eq \ref{eq:kappa}) with the determined $\widetilde{C_{V}}$ ($\approx$ 1.8), we predicted the thermal conductivity in the whole $T$-$\rho$ range. Our results are shown  in Figure \ref{fig:kappa_pred}.
Based on the fitting error (Figure \ref{fig:model_fit} and eq \ref{eq:RE_model}), the average RE of our predicted $\kappa$ should be approximately 10 \% and  
the maximum RE  $\sim$ 20 \%. 
We note that due to finite-size effects, our prediction here may also be slightly under-estimated. 

Finally, by substituting $\delta = (\rho/m_{0})^{-1/3}$ into eq \ref{eq:kappa}, we obtain the relation $\kappa \propto \rho^{2/3} C_{S}(T, \rho)$.
As shown in Figure \ref{fig:EOS_HPT}C, $C_{S}$ increases slightly with $T$ but significantly with $\rho$, which, according to the model (eq \ref{eq:kappa}), leads to the same dependence found in our simulations for $\kappa$.



\begin{figure*}[!htp]
\centering
  \includegraphics[width=0.6\textwidth]{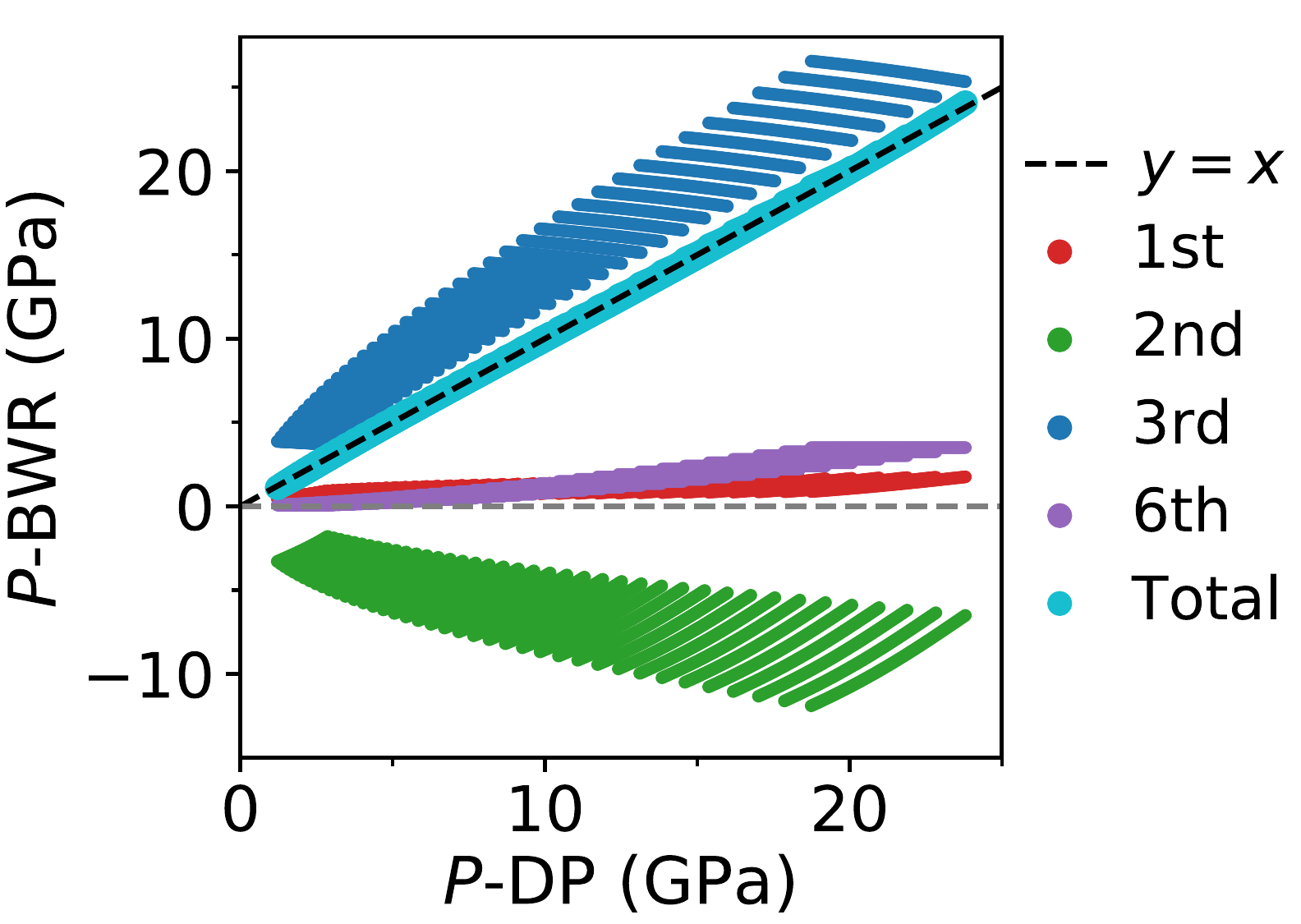}
  \caption{ 
  Pressure  of water obtained with the  Benedict–Webb–Rubin (BWR) equation ($P$-BWR) as a function of pressure computed in our simulations ($P$-DP).  We show values for several $T$ and $\rho$ conditions.
  The contributions from virial terms (1st, 2nd, 3rd and 6th) to $P$-BWR  are also shown.
  We plot a dashed line, $y=x$ as a guide to the eye, showing that $P$-BWR and $P$-DP values are close to each other.
  The $P$-BWR and $P$-DP are obtained on a dense grid, and the parameters in the BWR model are optimized (see text).
  }
  \label{fig:EOS_BWR}
\end{figure*}

We note that $C_{S}$ is related to the derivative of $P$ (see Methods); hence  an analytical formula for $P(T, \rho)$ is desirable to derive the relation between $C_{S}$ and $P$.
To this end, we fit our interpolated function $P(T, \rho)$ using the Benedict–Webb–Rubin (BWR) equation \cite{benedict1940_BWR,starling1973_BWR}:
\begin{equation}
\label{eq:BWR}
P(T, \rho) = \frac{ k_{\rm B}T }{ m_{0} } \rho + (B_{0}T + B_{1}) \rho^{2}
+ (C_{0}T + C_{1}) \rho^{3} + D \rho^{6}
\end{equation}
where $m_{0}$ is the mass of a water molecule; $B_{0}$, $B_{1}$, $C_{0}$, $C_{1}$ and $D$ are fitting parameters. We have ignored the exponential term and terms higher than $(1/T)^{0}$ for simplicity. 
Our fitting data, denoted as $P$-DP, are evenly spaced over a 100 $\times$ 40 mesh;  at each grid point $T$-$\rho$, a value of $P$ is obtained from the interpolated function $P(T, \rho)$. 
We optimized the parameters entering the BWR equation and we show the computed pressure ($P$-BWR) in Figure \ref{fig:EOS_BWR}, as well as the respective contributions.
Interestingly, the BWR equation accurately describes the interpolated function $P(T, \rho)$, with a small root-mean-square-error of $\sim$ 0.07 GPa. We find that the 3rd-term is dominant; the contributions of 1st-, 2nd- and 6th-terms are smaller than that of the cubic one. 
We note an approximate cancellation between the sum of the positive 1st- and 6th-terms and the negative 2nd-term; the 3rd-term alone is of similar magnitude to the total pressure ($P$-BWR). Hence, for simplicity, we  assume $P(T, \rho) \sim C(T) \rho^{3}$, where $C(T)$ refers to $C_{0}T + C_{1}$ in eq \ref{eq:BWR}. 

Knowing the dependence of the sound velocity on pressure and a form of the pressure as a function of temperature and density, we can now obtain an approximate dependence of $\kappa$ on the pressure:
\begin{equation}
\label{eq:sqrt_derive}
\kappa \propto C_{S}/\delta^{2} \propto \rho^{2/3} 
\sqrt{ \gamma \Bigl( \frac{\partial P} {\partial \rho} \Bigr)_{T} }
\sim \rho^{1/6} \sqrt{ \gamma \times 3 C(T) \rho^{3} }
\propto \rho^{1/6} \sqrt{ \gamma } \sqrt{ P } 
\end{equation}  
where $\gamma$ is the adiabatic index (see Methods). In the $T$-$\rho$ range studied here, we find that $\sqrt{\gamma}$ and $\rho^{1/6}$ are in the range of $\sim$ (1.0, 1.1), i.e. nearly constant; as a result, we obtain that $\kappa \propto \sqrt{P}$. 
Although the square-root relation $\kappa \propto \sqrt{P}$ found here is not rigorously  proven, it is  a simple and useful functional relationship to approximately predict $\kappa$ when $P$ is measured in the range investigated in our work.

\section{Conclusions}

By carrying out  SAEMD simulations with the DP potential, we computed the thermal conductivity of  water at high temperatures, 1,000 $< T <$ 2,000 K and 1.0 $< \rho <$ 1.86 g/cm$^3$, at conditions relevant to the Earth mantle.
We found that the thermal conductivity depends weakly on the temperature and increases  monotonically with the density and pressure, reaching values approximately 4 times larger than that at ambient conditions at the highest density point, indicating a more efficient heat energy transport under pressure than at ambient conditions.
We showed that a simple  model (eq \ref{eq:kappa}) can satisfactorily describe the thermal conductivity of water at extreme conditions and using such a model we provided predictions of the thermal conductivity in a broad range of density and temperature. 
Our results indicate that the heat is transferred roughly at the speed of sound over nearest-neighbors  inter-molecular distances, and that the heat conduction mechanism is not significantly affected by water dissociation, when the proportion of dissociated molecules remain smaller than 15 $\%$. Our simulations and the model used here to interpret them indicate that an increased sound velocity and density at extreme conditions are responsible for a larger thermal conductivity in HPT water than at ambient conditions.
Numerically, we identified a square-root relationship between the thermal conductivity and the pressure of the system. 
Although this relationship  is not rigorous, it can be useful to estimate the thermal conductivity at $T$-$\rho$ conditions similar to those studied here, since its direct measurement may be more difficult than that of the pressure.
Our study provides both insights and useful data on transport properties and equation of states of water at high temperature and pressure, which may be useful in planetary and geo-sciences.


\section*{Notes}
The authors declare no competing financial interest.

\section*{Acknowledgements}
The work of CZ, MP and GG was supported by MICCoM, as part of the Computational Materials Sciences Program funded by the U.S. Department of Energy, Office of Science, Basic Energy Sciences, Materials Sciences, and Engineering Division through Argonne National Laboratory. The work of LZ and RC was supported by the Computational Chemical Sciences Center “Chemistry in Solution and at Interfaces” funded by the U.S. Department of Energy under Award No. DE-SC0019394.
We acknowledge the computational resources at the University of Chicago's Research Computing Center and the Princeton Research Computing resources at Princeton University.

\bibliography{Ref_main}


\end{document}